\documentclass[fleqn,twoside]{article}
\usepackage{espcrc2}
\usepackage{graphicx}
\usepackage[figuresright]{rotating}
\usepackage{lineno}

\newcommand{\nubar}{\bar{\nu}}
\newcommand{\numu}{\nu_{\mu}}
\newcommand{\numubar}{\bar{\nu}_{\mu}}
\newcommand{\nue}{\nu_e}
\newcommand{\nuebar}{\bar{\nu}_e}
\newcommand{\mutoe}{\numu\rightarrow\nue}
\newcommand{\mubartoebar}{\numubar\rightarrow\nuebar}
\def\np#1#2#3   {{ Nucl. Phys.} {\bf#1}, #2 (#3)}
\def\pl#1#2#3   {{ Phys. Lett.} {\bf#1}, #2 (#3)}
\def\prev#1#2#3 {{ Phys. Rev.} {\bf#1}, #2 (#3)}
\def\prd#1#2#3 {{ Phys. Rev.} {\bf D #1}, #2 (#3)}

\title{Prospects for Antineutrino Running at MiniBooNE}

\author{M.O. Wascko\address{Department of Physics and Astronomy,\\
                 Louisiana State University, Baton Rouge, LA 70803},
                  for the MiniBooNE Collaboration}
       
\begin{document}

\begin{abstract}
  MiniBooNE began running in antineutrino mode on 19 January, 2006.
  We describe the sensitivity of MiniBooNE to LSND-like $\nuebar$
  oscillations and outline a program of antineutrino cross-section
  measurements necessary for the next generation of neutrino
  oscillation experiments.  We describe three independent methods of
  constraining wrong-sign (neutrino) backgrounds in an antineutrino
  beam, and their application to the MiniBooNE antineutrino analyses.
\end{abstract}

\maketitle

\section{Introduction}

MiniBooNE~\cite{boone-prop} is a neutrino oscillation experiment at
Fermilab, designed to confirm or rule out the hypothesis that the LSND
$\nuebar$ excess~\cite{lsnd} is due to $\mubartoebar$ oscillations. A
general description of the experiment can be found
elsewhere~\cite{runplan}.  Heretofore, MiniBooNE has been taking data
in neutrino mode, searching for $\mutoe$ oscillations.  However, in
some scenarios involving CP and CPT violation, oscillations may occur
only in antineutrinos.  Thus, searching for oscillations in
antineutrinos is a crucial test~\cite{aps}.

Furthermore, the search for CP violation in the neutrino sector by
future off-axis experiments~\cite{t2k-prop,nova-prop} requires
$\mubartoebar$ oscillation measurements, as well as $\mutoe$.  The
signature for CP violation is an asymmetry in these oscillation
probabilities, but this can only be confirmed if the precision of the
$\nu$ and $\nubar$ cross sections are smaller than the observed
asymmetry.  There are few $\nu$ cross section data
published~\cite{xsec-nu} to date, but even fewer measurements of low
energy $\nubar$ cross sections.  We will need more and better data if
we hope to find CP violation in the neutrino sector.

\begin{table}[h]
  \caption{\em Event rates expected in MiniBooNE $\nubar$ running with 
    $2\times10^{20}$ POT assuming a 550 cm fiducial volume,  
    before cuts. Listed are the expected right-sign (RS)
    and wrong-sign (WS) events for each reaction channel.}
   \label{table:nubar-event-stats}
\centering
   \begin{tabular}{ c c c }
\\Reaction        & $\numubar$ (RS)  & $\numu$ (WS)  \\ \hline
  CC QE                    &  32,476 &  11,234  \\ 
  NC elastic               &  13,329 &   4,653  \\ 
  CC resonant $1\pi^-$     &   7,413 &      0   \\ 
  CC resonant $1\pi^+$     &       0 &   6,998  \\ 
  CC resonant $1\pi^0$     &   2,329 &   1,380  \\ 
  NC resonant $1\pi^0 $    &   3,781 &   1,758  \\ 
  NC resonant $1\pi^+$     &   1,414 &     654  \\ 
  NC resonant $1\pi^-$     &   1,012 &     520  \\ 
  NC coherent $1\pi^0$     &   2,718 &     438  \\ 
  CC coherent $1\pi^-$     &   4,487 &       0  \\ 
  CC coherent $1\pi^+$     &       0 &     748  \\ 
  other (multi-$\pi$, DIS) &   2,589 &   2,156  \\ \hline
  total                    &  71,547 &  30,539  \\ \hline   
   \end{tabular}
\end{table}

\begin{figure}[htb]
\center
{\includegraphics[width=3.0in]{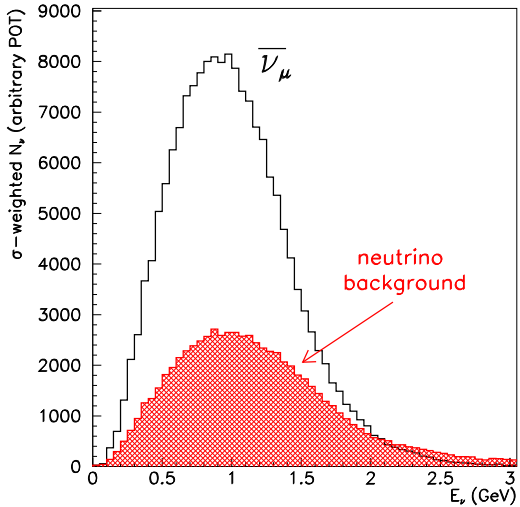}}
\vspace{-0.5in}
\caption{\em Comparison of cross section weighted right sign (black
  histogram) and wrong sign (red cross-hatched histogram) fluxes in
  antineutrino mode.}
\label{fig:ws_nubar}
\end{figure}

Table~\ref{table:nubar-event-stats} lists the expected antineutrino
event statistics for $2\times10^{20}$ protons on target (POT) of
$\nubar$ running with MiniBooNE~\cite{runplan,loi}, based on the
\texttt{nuance} Monte Carlo~\cite{nuance}.  Rates are listed for both
right-sign (RS) and wrong-sign (WS) interactions.  Note that
wrong-signs comprise $\sim~30\%$ of the total events in antineutrino
mode, as illustrated in Figure~\ref{fig:ws_nubar}.  To constrain the
wrong-sign backgrounds, MiniBooNE has developed several new analysis
techniques. We describe three methods below and detail their
application to $\nubar$ cross section measurements at MiniBooNE.

%
%

\section{Constraining Wrong Sign Events}
\label{sec:ws-constraints}

\begin{figure}[t]
\center
{\includegraphics[width=3.0in]{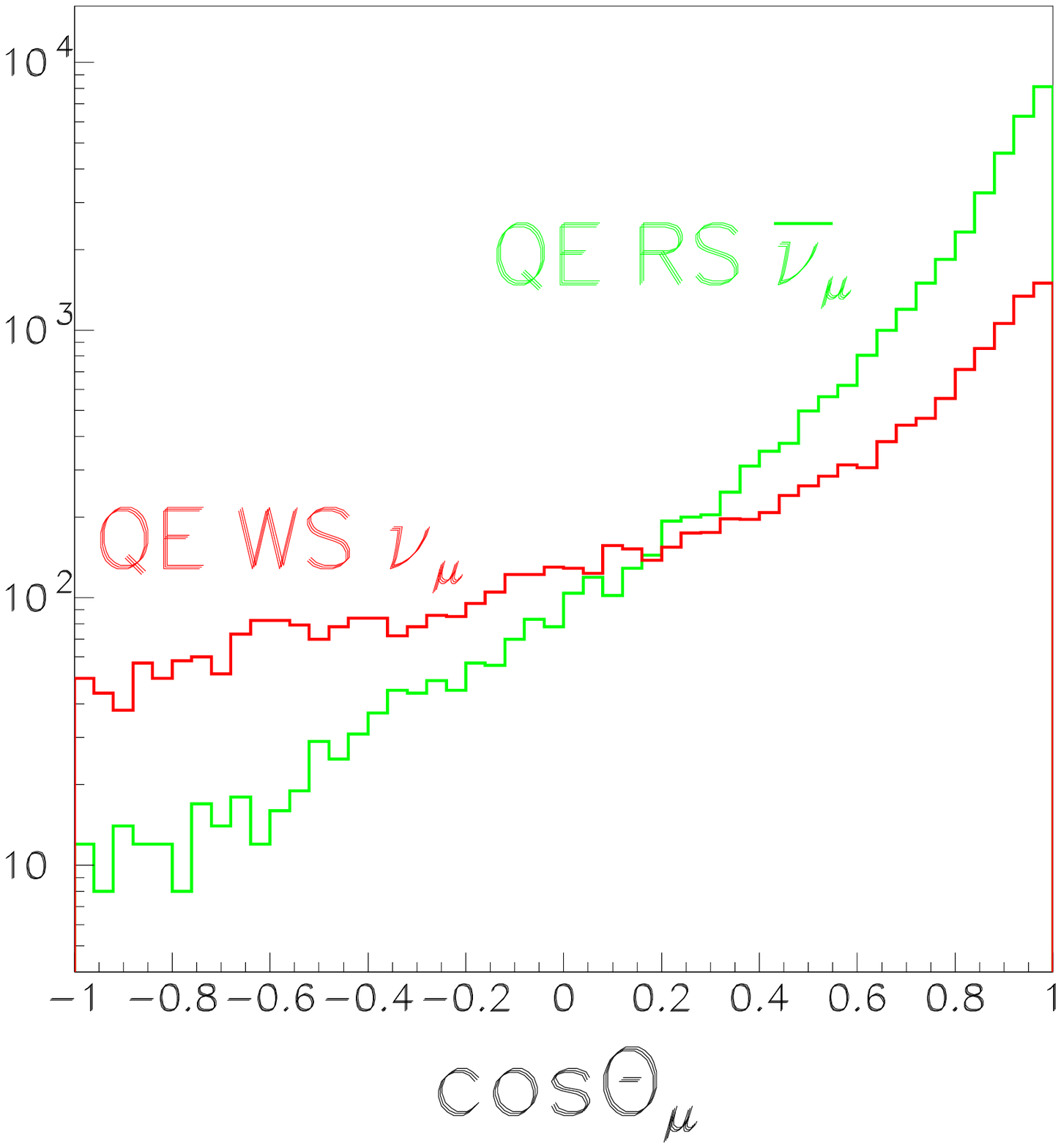}}
\vspace{-0.5in}
\caption{\em Generated muon angular distributions for CC QE right-sign
  $\numubar$ and wrong-sign $\numu$ interactions in antineutrino
  running at MiniBooNE. }
\label{fig:muang_gen}
\end{figure}

For charged current (CC) interactions, neutrino events are typically
distinguished from antineutrino events by identifying the charge of
the outgoing muon.  MiniBooNE, which has no magnetic field, cannot
distinguish $\nu$ from $\nubar$ interactions on an event-by-event
basis.  Instead, we have developed several novel techniques for
measuring WS backgrounds in antineutrino mode data; this will allow
more precise $\nubar$ cross section measurements. The WS content is
constrained by three measurements: muon angular distributions in
quasi-elastic (CC QE) events; muon lifetimes; and the measured rate of
CC single charged pion (CC1$\pi^+$) events~\cite{loi}.  Combined,
these three independent measurements (each with different systematic
uncertainties) offer a very powerful constraint on the neutrino
backgrounds in antineutrino mode (Table~\ref{table:ws-uncertainties}).

\subsection{Muon Angular Distributions}

The most powerful wrong-sign constraint comes from the observed
direction of outgoing muons in CC QE interactions. Neutrino and
antineutrino events exhibit distinct muon angular distributions.  Due
to the antineutrino helicity, the distribution of final state muons in
$\numubar$ QE events is more forward peaked than that from $\numu$
interactions.  This is illustrated in Figure~\ref{fig:muang_gen}, which
shows the angle of the out-going lepton with respect to the neutrino
beam axis for both $\numu$ and $\numubar$ CC QE events from the
\texttt{nuance} Monte Carlo~\cite{nuance}.

\begin{figure}[t]
\center
{\includegraphics[width=3.0in]{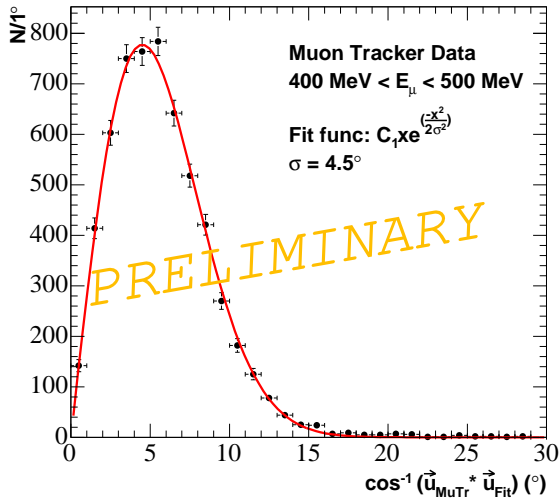}}
\vspace{-0.5in}
\caption{\em Angular resolution of cosmic muons in MiniBooNE measured
  using the muon tracker calibration system. The plot shows the angle
  between the reconstructed muon direction from the tank event fitter
  and the direction from the muon tracker.  This plot shows only
  events with reconstructed kinetic energy between 400 and 500 MeV,
  pointed into the fiducial volume of the tank. The fitted width of
  the distribution is 4.5 degrees, and the intrinsic resolution of the
  muon tracker is ~2 degrees. Assuming the resolutions add in
  quadrature, this yields an angular resolution of 4 degrees for the
  tank event fitter.}
\label{fig:mutr_delmt}
\end{figure}
\begin{figure}[t]
\center
{\includegraphics[width=3.0in]{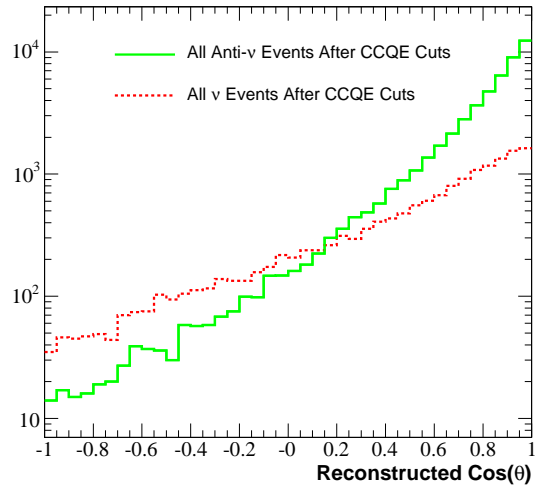}}
\vspace{-0.5in}
\caption{\em Reconstructed muon angular distributions for CC QE
  right-sign $\numubar$ and wrong-sign $\numu$ interactions in
  antineutrino mode at MiniBooNE. }
\label{fig:muang_recon}
\end{figure}
\begin{table}
   \caption{\em Wrong-sign extraction uncertainties as obtained from 
            various independent sources in the $\nubar$ data. The
            resultant systematic uncertainty on $\nubar$ cross section
            measurements is obtained by assuming that wrong-signs
            comprise $30\%$ of the total events.}
   \label{table:ws-uncertainties}
\centering
   \begin{tabular}{ c c c }
\\Measurement  &   \hspace{0.2cm}WS \hspace{0.2cm} &  \hspace{0.2cm}resultant \\
                           &   \hspace{0.2cm}uncertainty\hspace{0.2cm} & \hspace{0.2cm} error on $\sigma_{\nubar}$ 
\\ \hline
  CC QE $\cos\theta_\mu$   &  $7\%$   &  $2\%$   \\
  CC $1\pi^+$ cuts         &  $15\%$  &  $5\%$   \\
  muon lifetimes           &  $30\%$  &  $9\%$   \\ \hline
   \end{tabular}
\end{table}

In order to use this technique to constrain the WS background, the
angular resolution for muons must be sufficiently good to separate the
two populations. MiniBooNE's angular resolution is measured using the
cosmic muon calibration system, which consists of a muon tracker
hodoscope placed above the detector.  Figure~\ref{fig:mutr_delmt}
shows the 4$^{\circ}$ angular resolution of muons between 400 and 500
MeV.  This resolution is sufficient to allow exploitation of the
angular differences in the CCQE outgoing muons distributions.  The
angular distributions can be fitted to extract the wrong-sign
contribution.  Analysis of our Monte Carlo data sets indicates that
the wrong-sign content can be measured using this technique, with a
statistical uncertainty of 5\% of itself~\cite{hiro-memo}.  This is
illustrated in Figure~\ref{fig:muang_recon}, which shows the
reconstructed angle of the out-going lepton with respect to the
neutrino beam axis for both $\numu$ and $\numubar$ CC QE events
generated with the MiniBooNE Monte Carlo, after all selection cuts
have been applied.  Including systematic uncertainties and (non-QE)
backgrounds increases the uncertainty on the WS extraction only to
$7\%$.

\subsection{Muon Lifetimes}

A second constraint results from measuring the rate at which muons
decay in the MiniBooNE detector. Due to an $8\%$ $\mu^-$ capture
probability in mineral oil, negatively and positively charged muons
exhibit different effective lifetimes ($\tau = 2.026 \, \mu s$ for
$\mu^-$~\cite{mu-lifetime-cap} and $\tau = 2.197 \, \mu s$ for
$\mu^+$~\cite{mu-lifetime-std}).  To extract the WS fraction, one
simply fits the muon lifetime with a sum of two exponentials, with the
extracted fraction of each exponential term giving the fraction of RS
or WS events.  For CCQE events, we find that the wrong-sign
contribution can be extracted with a $30\%$ statistical uncertainty
based solely on this lifetime difference and negligible systematic
uncertainties. While not as precise as fits to the muon angular
distributions, this particular constraint is unique, as it is
independent of kinematics.

\subsection{CC Single Pion Event Sample}

Our third wrong-sign constraint employs the the fact that
antineutrinos do not create CC$1\pi^+$ events in the detector---these
all stem from neutrinos
(Table~\ref{table:nubar-event-stats}). MiniBooNE identifies CC$1\pi^+$
events by tagging the two decay electrons that follow the primary
neutrino interaction, one from the $\mu^-$ and one from the $\pi^+$
decay~\cite{morgan-ccpip}.  However, CC$1\pi^-$ events do not pass
this requirement because all the emitted $\pi^-$'s which stop in the
detector descend into atomic orbits around carbon nuclei and are
instantly captured, leaving no decay electrons~\cite{lsnd-nim}.  The
$\pi^-$ decay in flight rate is much smaller than the rate of $\pi^+$
decays at rest~\cite{lsnd-nim}.  Thus, applying the two Michel cut to
the full sample, which is $70\%$ antineutrino (RS) interactions,
yields an $85\%$ pure sample of WS neutrino events.

Assuming conservative uncertainties for the antineutrino background
events and the CC$1\pi^+$ cross section, which is currently being
measured by MiniBooNE, we expect a $15\%$ uncertainty on the
wrong-sign content in the beam given $2\times10^{20}$ POT.  This
constraint is complementary to the muon angular distributions, because
CC$1\pi^+$ events stem mainly from resonance decays, thus constraining
the wrong-sign content at larger neutrino energies.

%

\section{CC Quasi-Elastic Scattering}

MiniBooNE expects more than 40,000 QE interactions in antineutrino
mode with $2\times10^{20}$ POT before cuts. Using the same QE event
selection criteria as the previously reported MiniBooNE neutrino
analysis~\cite{jocelyn-ccqe} yields a sample of $\sim19,000$ events,
with $75\%$ QE purity with both WS and RS events.

Assuming the above wrong-sign constraints and conservative errors on
the $\nu$ flux, backgrounds, and event detection, we expect a
MiniBooNE measurement of the $\numubar$ QE cross section to better
than $20\%$ with $2\times10^{20}$ POT.

\section{NC Single Pion Production}

$\numubar$ neutral current (NC) $\pi^0$ production is one of the largest backgrounds to
future $\mubartoebar$ oscillation searches. There has been only one
published measurement of the absolute rate of this channel, with
$25\%$ uncertainty at 2 GeV~\cite{ncpi0-faissner}.

\begin{figure}[t]
\center
{\includegraphics[width=3.0in]{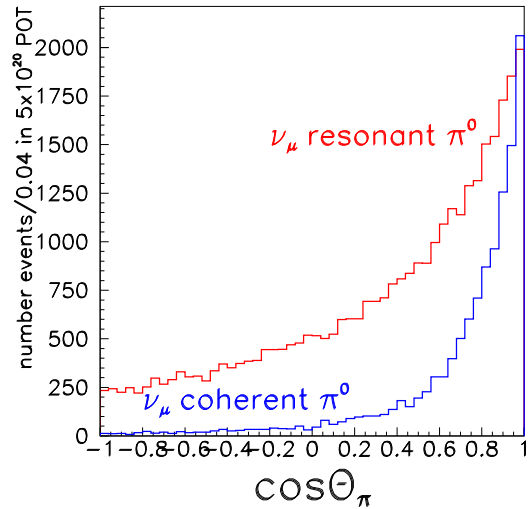}}
\vspace{-0.5in}
\caption{\em Generated $\pi^0$ angular distributions for NC $\nu$
  scattering~\cite{nuance}. This is the angle of the outgoing $\pi^0$
  in the lab with respect to the $\nu$ direction.}
\label{fig:ncpi0-nu}
\end{figure}

\begin{figure}[t]
\center
{\includegraphics[width=3.0in]{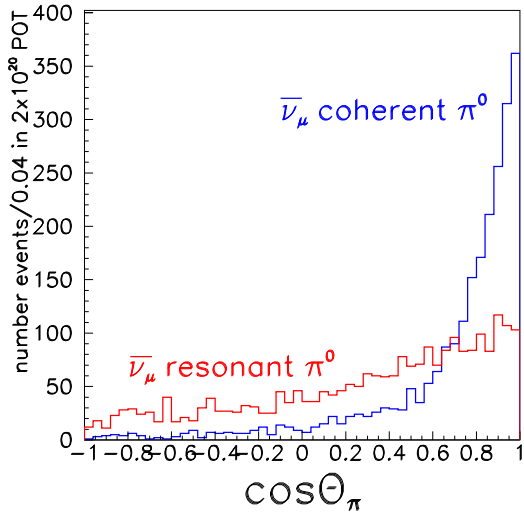}}
\vspace{-0.5in}
\caption{\em Generated $\pi^0$ angular distributions for NC $\nubar$
  scattering~\cite{nuance}. This is the angle of the outgoing $\pi^0$
  in the lab with respect to the $\nu$ direction. The coherently
  produced $\pi^0$'s are more forward peaked than the resonant
  $\pi^0$'s in both cases.}
\label{fig:ncpi0-nubar}
\end{figure}

Applying MiniBooNE's $\numu$ NC $\pi^0$ cuts~\cite{jen-pi0} with no
modifications leaves a sample of antineutrino NC $\pi^0$ events with a
similar event purity and efficiency. After this selection, we expect
1,650 $\numubar$ resonant NC $\pi^0$ events and 1,640 $\numubar$
coherent NC $\pi^0$ events assuming $2\times10^{20}$ POT and the Rein
and Sehgal model of coherent pion
production~\cite{nuance,rein-sehgal-coh}. Coherent pion production has
a characteristic pion angular distribution that allows it to be
distinguished from resonant production, as illustrated in
Figures~\ref{fig:ncpi0-nu} and \ref{fig:ncpi0-nubar}.  Moreover, the
previously mentioned figures illustrate that the distinctiveness of
the angular distributions should be even more marked in antineutrino
running, which increases the value of antineutrino data for
understanding coherent production.  The background of $\sim$~1000 WS
events will be determined by the constraints on the wrong-sign content
in the beam as described in Section~\ref{sec:ws-constraints} and the
measurement of the $\numu$ NC $\pi^0$ cross section from MiniBooNE
neutrino data.

\section{CC Single Pion(CC1$\pi^-$) Production}

MiniBooNE expects roughly 7,000 resonant CC $1\pi^-$ with
$2\times10^{20}$ POT before cuts. As discussed above, almost all of
the emitted $\pi^-$'s will be absorbed by carbon nuclei, and will
therefore not be selected by the CC1$\pi^+$ cuts. Nevertheless, these
events still have a signature: two Cherenkov rings (one each from the
$\mu^+$ and $\pi^-$) and one Michel electron in the vicinity of the
$\mu^-$. The selection efficiency and purity of such events is unknown
at this time.  Further investigation is currently underway.

\section{Oscillation Sensitivity}

Although MiniBooNE has been searching for oscillations in neutrino
mode, the LSND oscillation signal was actually an excess of
antineutrino events.  Because of the potential for finding CP
violation in the neutrino sector, it is imperative that MiniBooNE test
the LSND oscillation hypothesis with antineutrino data~\cite{aps}.

\begin{figure}[t]
\center
{\includegraphics[width=3.0in]{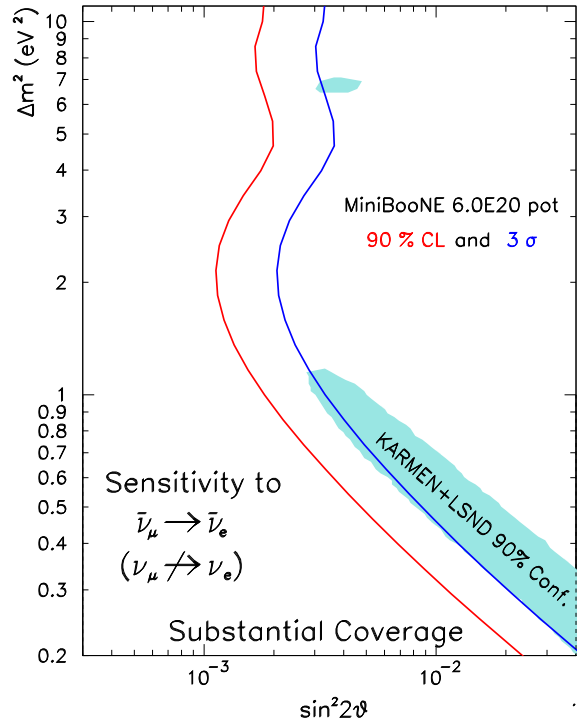}}
\vspace{-0.5in}
\caption{\em \small Sensitivity to $\numubar\!\rightarrow\!\nuebar$ 
oscillations assuming no $\numu\!\rightarrow\!\nue $ oscillations using 
an energy spectrum fit. Shown in blue is the allowed 90\% CL region from a 
joint analysis of the KARMEN and LSND $\numubar\!\rightarrow\!\nuebar$
oscillations results~\cite{joint-karmen-lsnd}.}
\label{fig:nuebar-app}
\end{figure}

Figure~\ref{fig:nuebar-app} shows an estimate of the MiniBooNE
sensitivity to $\mubartoebar$ oscillations under the assumption that
$\mutoe$ oscillations do not occur, i.e. assuming MiniBooNE sees no
$\nue$ appearance signal.  Here, we compare the sensitivity to the
joint KARMEN-LSND region ($\nubar$ only)~\cite{joint-karmen-lsnd}, not
the full LSND allowed region ($\nu + \nubar$).  This is a
statistics-limited search; further running is needed beyond the
$2\times10^{20}$ POT assumed in the preceding sections in order to
test the LSND hypothesis.  This sensitivity shown assumes
$6\times10^{20}$ POT.

%

\section{Conclusions}

We have developed three techniques for determining the wrong-sign
background in antineutrino mode.  The resulting systematic error on
any given $\nubar$ cross section measurement due to the wrong sign
contamination should be less than 2\% averaged over the entire flux,
which is remarkable for a detector which does not possess
event-by-event sign selection. Given this redundant approach, the
wrong-sign contamination should not be considered prohibitive to
producing meaningful antineutrino cross section~\cite{loi} and
oscillation measurements~\cite{loi,jocelyn-memo,alexis-memo} at
MiniBooNE. These techniques may also be useful for other experiments
without magnetized detectors which have plans to study antineutrino
interactions ({\em e.g.} T2K, NO$\nu$A, Super-K).

MiniBooNE began running in antineutrino mode on 19 January, 2006, and
is currently approved to run for one more year.  In order to truly
confirm or rule out the LSND oscillation hypothesis, MiniBooNE needs
6$\times 10^{20}$ POT, which will require additional years of running.

\section{Acknowledgments}

The author is pleased to acknowledge the collaborative efforts of J.M.
Link, H.A. Tanaka, and G.P. Zeller in developing the ideas in this
work.  The author would also like to express gratitude to the
organizers of NuInt05 for their generous travel support.

The MiniBooNE collaboration gratefully acknowledges support from
various grants and contracts from the Department of Energy and the
National Science Foundation. The author was supported by grant number
DE-FG02-91ER0617 from the Department of Energy.


\end{document}